\documentclass[11pt]{amsart}
\usepackage{geometry}                % See geometry.pdf to learn the layout options. There are lots.
\usepackage{graphicx}
\usepackage{amssymb}
\usepackage{epstopdf}
\newcommand{\eps}{\varepsilon}
\newcommand{\R}{\mathbb{R}}

\title{Entropy for a class of micro-economic models}
\author{R.S.MacKay}
\address{Mathematics Institute and Centre for Complexity Science, University of Warwick, Coventry CV4 7AL, UK}
\email{R.S.MacKay@warwick.ac.uk}
\date{\today}                                           % Activate to display a given date or no date

\begin{document}

\begin{abstract}
Chater and MacKay \cite{CM} derived an entropy function of state for exchange economies satisfying a list of axioms, and showed that a change of state of a system of such economies is possible if and only if their total entropy does not decrease.  In this paper, a large class of agent-based models is proved to satisfy the axioms in the thermodynamic limit, and the entropy is shown to be the logarithm of the partition function for their stationary distributions.
\end{abstract}

\maketitle

\section{Introduction}
\label{sec:intro}

This paper considers exchange economies in a ``thermodynamic'' limit.  In exchange economies, agents exchange amounts of various types of durable good \cite{GG,TSB}.  The thermodynamic limit is to take the number of agents to infinity with the mean amounts of goods per agent going to finite positive limits and some asymptotic scaling assumptions on the dynamics.

The main models to be considered, called here {\em basic economies}, have a large number $N$ of agents, and a finite number $L$ of types of infinitely divisible durable good including one called money.  Each agent $i$ has a positive ``utility'' function $u_i$ of its non-negative vector $p_i \in \R_+^L$ of amounts of possessions and possibly of the holdings of others.  Agents $i$ and $j$ make pairwise encounters independently at some rates $k_{ij}$ (symmetric), such that the encounter graph is connected.  %We require the connectivity to be uniform in $N$ [NOT YET USED!], specifically so that there is a $\lambda>0$ independent of $N$ such that for all $v \in \R^N$ with $\sum_i v_i = 0$ then $v^Tkv \le \sum_i (K_i-\lambda) v_i^2$, where $K_i = \sum_j k_{ij}$ (so that the first eigenvalue of the graph Laplacian is positive).  
On encounter, the pair pool their possessions and redistribute them between the two with probability density proportional to the product of their utilities for the outcome\footnote{or they could do this for a pre-agreed fraction of their possessions.}.

The resulting dynamics define a reversible Markov process with respect to a stationary probability distribution that has ``density''  (in quotes because a delta-function is used to represent the restriction to a subspace)
$$\rho(p) = \frac{1}{Z(P)} \prod_i u_i(p)\, \delta(\sum_i p_i - P)$$ with respect to $\prod_i dp_i$, where $dp_i$ denotes the standard measure $\prod_{t=1}^L dp^t$ on $\R_+^L$ for agent $i$, $P \in \R_+^L$ is the vector of totals of the types of good, $\delta$ is the Dirac delta-function (in $L$ dimensions), and 
\begin{equation}
Z(P) = \int \prod_i u_i(p)\, \delta(\sum_i p_i - P) \, \prod_i dp_i
\label{eq:Z}
\end{equation}
is a normalisation constant called the {\em partition function}.  Reversibility with respect to $\rho$ is easy to check. The stationary probability $\rho$ is assumed to attract all initial probabilities in total variation metric.  This is proved for all-to-all encounters ($k_{ij}>0$ for all $i\ne j$) of ``Cobb-Douglas'' agents\footnote{whose utilities depend only on their own possessions $p$ and by a power law $u_i(p) \propto \prod_t p_t^{\alpha_t-1}$, where $t$ labels type of good, for some exponents $\alpha_t$ that may depend on $i$ also.}  in \cite{M25} and is expected to hold more generally.

The basic model can be generalised to {\em multi-part economies}, for which there is a finite partition of the agents into sub-economies (parts) such that for each pair of parts, only certain types of good can be exchanged between them.  Thus the outcome of an encounter between a pair of agents from different parts is given by pooling and redistributing only the specified types of good.  As a result there are in general additional  conserved quantities beyond the amounts of types of good, namely the amount of some good in a union of parts which that good can not leave or enter.  
An obvious example is an economic system consisting of two unconnected basic economies, but more general examples arise by for example allowing money to flow between two parts but not goods.
Supposing that one has identified a complete independent set of such conserved quantities, there is a stationary probability density of the same form as above but where $P$ is generalised to the amounts of the conserved quantities.  For example, if the economy can be partitioned into two parts $A$ and $B$, has money $M$ and one type of good $G$, and money can flow between $A$ and $B$ but the good can not, then the constraint in the stationary density becomes 
$$\delta(\sum_{i \in A\cup B} m_i - M) \delta(\sum_{i\in A} g_i - G_A) \delta(\sum_{i \in B} g_i - G_B),$$ 
and the partition function becomes a function of $(M,G_A,G_B)$.

An economy is called {\em simple} if it has money as one good and if one of the connected components for flow of money is distinguished.  
Then we use $M$ for the amount of money in the distinguished component (in the case where there is only one financial component, then $M$ is the total money).  Attention will be focussed on simple economies, but the goal is to determine what can happen on putting one or more of them into contact.  Subject to the assumptions of \cite{CM}, the answer is that changes can occur if and only if the total ``entropy'' does not decrease, a function of the state for each simple system, obtained by a theorem of \cite{LY}.

The main aim of this paper is to show that under assumptions to be stated in the next section, such models satisfy the conditions for the thermal macroeconomic theory of \cite{CM} in the thermodynamic limit, and to compute the resulting economic entropy functions of state.  The result is that $S=\log Z$ (up to adding an arbitrary multiple of $N$ and an overall scale factor).  

An outline of the paper is presented at the end of the next Section, after explaining the strategy and describing the assumptions.

%As a subsidiary result, it is shown how to compute $S$ from the ``canonical'' partition function $Z_c$, which in some models is easier to compute than $Z$.

\section{Strategy and Assumptions}
\label{sec:strategy}
To summarise the theory of \cite{CM}, from some axioms for macro-economic systems, it is deduced by application of the theory of \cite{LY} that there is an entropy function of system state such that any change in the states of a collection of economies put into contact does not decrease the total entropy.  This puts flesh on the idea of ``gains of trade'' but leads to much more, such as a notion of economic temperature governing flow of money, well defined concepts of value of money and goods, and so of inflation and market prices, and non-positivity and symmetry of various partial derivatives of values of goods with respect to amounts of goods in the economy.

Formally, the micro-economic systems introduced here do not fit in the framework of our thermal macroeconomic theory, because one of the required axioms (A4) is that for any system $A$ and $\lambda>0$ one can consider a scaled version $\lambda A$, but if $A$ has $N$ agents then $\lambda A$ would in general have non-integer number ($\lambda N$) of agents.  Nonetheless, we can consider sequences of these systems with $N$ going to infinity and require ``extensivity'' assumptions on the way the sequence behaves as $N \to \infty$ and consider sequences of allowed scaling factors.  
This is the concept of the ``thermodynamic limit'' from physics.

Specifically, the amounts of each type of good (and conserved quantities more generally) are taken to be asymptotically proportional to the number $N$ of agents.  The utility functions for the agents are drawn from a common distribution. 
The sizes of the partitions in a multi-part economy are taken asymptotically proportional to $N$.  As a result, from (\ref{eq:Z}), $\log Z$ is asymptotically proportional to $N$ and can be written as 
\begin{equation}
\log Z(P) = N \zeta(P/N)
\label{eq:zeta}
\end{equation}
for a $C^{1+Lip}$ function\footnote{$\zeta \in C^{1+Lip}$ means that $\zeta$ is differentiable and its derivative is Lipschitz continuous.} $\zeta$ of the vector $p = P/N$ of mean possessions per agent (that may also depend on $N$ but has a limit as $N \to \infty$).  The quantity $\zeta(p)$ will turn out to be the entropy per agent in the macroeconomic theory.

Various further assumptions will be required.  Firstly, it is assumed that
\begin{equation}
\beta = \frac{\partial \zeta}{\partial m}  > 0,
\label{eq:beta}
\end{equation}
uniformly in $N$, where $m$ is the mean money per agent.
Secondly, for fixed amounts of all other goods, $\beta$ goes continuously from $+\infty$ to $0$ as $m$ goes from $0$ to $+\infty$ with moreover, $\frac{\partial \beta}{\partial m} < 0$.  Thirdly, we need there to be a choice of good different from money such that $\nu/\beta$ increases as $m$ increases, where $\nu = \frac{\partial \zeta}{\partial g}$, with $g$ the mean amount of that good per agent.

Quantities scaling like $N$ in the thermodynamic limit, like the vector $P$ of total amounts of possessions, are called {\em extensive}.  Quantities asymptotically independent of $N$, like $\beta$ in (\ref{eq:beta}), are called {\em intensive}.  We assume fourthly that intensive quantities vary like $1/N$ with respect to changes in extensive quantities.  We will also assume that all the first partial derivatives of $\zeta$ are positive, though this might not really be necessary.

To complete our assumptions, we have to specify how such an economy interacts wtih an idealised trader with unlimited assets and possible use of an external economy (to fit it in the framework of \cite{CM}), and make assumptions on how the utility of an agent depends on the possessions of others.

%We claim that, interpreted in this sense of thermodynamic limit, this class of models satisfies the Lieb\&Yngvason axioms \cite{LY} that we used in \cite{CM}, and that the resulting entropy function is $\log Z$ (up to arbitrary shift of origin for each economy and one overall scale factor).

%To prove these claims we have to specify how such an economy interacts with an idealised trader with unlimited assets.  

We write $X\preceq Y$ if the trader can move an economic system from state $X$ to $Y$ with possible change in the trader's assets and arbitrarily small change in the state of any external system that the trader may use (we will write $X\prec Y$ if $X\preceq Y$ and $Y \not\preceq X$).

One mode of interaction with the trader is financial contact.  The trader makes a pot of money available with an initial (extensive) amount $M_T$.  Agent $i$ encounters the trader at some rate $K_i$.  On encounter, agent $i$ pools their money $m_i$ with the present amount $m_T$ of money in the pot and the sum is redistributed between the two with probability density proportional to $u_i$ (the trader is assumed to have a flat utility function).  
%This is treated in Section~\ref{sec:fin}.

The trader can put two parts of an economy into contact, allowing exchange of a specified list of types of good.  %This is treated in Section~\ref{sec:making}.
Similarly, the trader can subdivide an economy into pieces with barriers for exchange of some types of good. % (Sec~\ref{sec:breaking}). 
For these, we will make a ``summability'' assumption on the dependence of agent's utility functions on the possessions of other agents, in Section~\ref{sec:making}, so that the effect of changing the set of other agents makes a negligible change at the aggregate level.

The final mode of interaction with the trader is trading contact. % (Sec.~\ref{sec:trade}).  
The trader posts a price vector $\mu_T$ (with all components positive) for one or more types of good other than money.  On encounter, agent $i$ updates their amounts of possessions with probability density proportional to its utility on the ``budget surface'' $m_i + \mu_T\cdot g_i$ constant.

One could consider other modes of interaction, but these four suffice for the theory.  It is important not to allow overly intricate modes of interaction.  Otherwise, as with Maxwell's demon in classical physics, one could end up with all states being accessible from all states and the theory becomes vacuous (entropy of any system is constant).

The paper shows firstly that, interpreted in this sense of thermodynamic limit, this class of models satisfies the Lieb\&Yngvason axioms \cite{LY} that we used in \cite{CM}.

The basic result of \cite{CM}, subject to the axioms, is that there is a function $S$ of state of economic systems such that one subdivided state $(X_k)_{k \in K}$ of an economy can be moved to another one $(X'_{k'})_{k' \in K'}$ by the trader if\footnote{Because the trader is allowed to leave the external system in an arbitrarily close state, the move can also be achieved in the case of no change in the total entropy.}
$$\sum_{k \in K}S(X_k) < \sum_{k' \in K'}S(X'_{k'}),$$ 
and cannot if 
$$\sum_{k \in K}S(X_k) > \sum_{k' \in K'}S(X'_{k'}).$$ 
The theory also shows that the function $S$ is unique up to addition of arbitrary constants for each economy and an overall positive scale factor.

The second result of the paper is that for our class of micro-economic models in the thermodynamic limit, one can take $S = \log Z$, with $Z$ being the partition function (\ref{eq:Z}).

The paper analyses the effects of the four modes of interaction with the trader (Sections~\ref{sec:fin}--\ref{sec:trade}), justifies the axioms for such models in Section~\ref{sec:checking}, and derives the formula $S = \log Z$ for the entropy in Section~\ref{sec:entropy}.  
Finally, an Appendix shows how the ``canonical ensemble'' can be used to simplify computation of the entropy in some examples.

\section{Financial contact with trader}
\label{sec:fin}

First, consider the effect of financial contact of a simple system with the trader.  
Denote by $M$ the initial amount of money in the distinguished financial component of the economy and $M_T$ the initial amount made available by the trader, and let $M' = M+M_T$.
The dynamics is again reversible, but with respect to a new equilibrium with probability density 
$$\frac{1}{\tilde{Z}(M',G)} \prod_i u_i\, \Theta(M'-\sum_i m_i)\delta(\sum_i g_i - G),$$ where the Heaviside function $\Theta(x) = 1$ for $x\ge 0$, $0$ for $x < 0$, $g_i$ denotes the vector of amounts of other types of good than money owned by agent $i$,  $G$ the vector of total amounts in the various parts of the economy, and 
$$\tilde{Z}(M',G) = \int \prod_i u_i\, \Theta(M'-\sum_i m_i) \delta(\sum_i g_i - G) \prod_i dm_i dg_i.$$  
Henceforth in this section, mention of dependence on $G$ is dropped, as $G$ is constant.  This equilibrium is assumed to be attracting.

It follows from (\ref{eq:Z}) that the marginal probability density for the amount $\tilde{M}=\sum_i m_i$ of money in the economy is $$Z(\tilde{M})\, \Theta(M'-\tilde{M})/\tilde{Z}(M').$$
%The normalisation constant $\tilde{Z} = \int \prod_i u_i \Theta(M+M_T-\sum_i m_i)\delta(\sum_i g_i - G)\, \prod_i dp_i$ can be written as
%$$\tilde{Z} = \int_0^{M+M_T} Z(\tilde{M}) d\tilde{M}.$$
%We see that the probability density for $\tilde{M}$ is $Z(\tilde{M}) \Theta(M+M_T - \tilde{M})/\tilde{Z}$.
%From the assumption (\ref{eq:beta}), $Z(\tilde{M}) = Z(M) \exp \int_M ^{\tilde{M}} \beta(M) dM$, so using Laplace's rule, $\tilde{Z}$ is to leading order $Z(M+M_T)/\beta(M+M_T)$.  Thus $\log{\tilde{Z}} \sim \log{Z(M)} + \int_M^{M+M_T} \beta(M) dM$, which exceeds $\log Z(M)$ by an extensive amount.  
On breaking the financial contact, the (distinguished component of the) economy will contain an amount $\tilde{M}$ of money with this probability density.  
From the definition (\ref{eq:beta}) of $\beta$, $$Z(\tilde{M}) = Z(M) \exp \int_M ^{\tilde{M}} \beta(M) dM.$$
Because $M_T$ is extensive and $\beta$ is assumed positive, the probability that $\tilde{M}\le M$ is exponentially small, so it follows that there was positive money flow to the economy.  In fact, virtually all the money $M_T$ flows to the economy, because for any $\eps>0$ the probability that $\tilde{M}<M+(1-\eps)M_T$ is exponentially small.  Denote the new state by $X{+M_T}$.

Also, assumption (\ref{eq:beta}) implies that 
$$\log Z(X{+M_T})  = \log Z(M) + \int_M^{M+M_T} \beta(\tilde{M})\, d\tilde{M} > \log Z(X).$$

So, for any positive (extensive) amount $M_T$ of money, the trader can move the state $X$ of any simple system to state $X{+M_T}$ and  $\log Z(X) < \log Z(X{+M_T})$.

\section{Making contact between parts of a system}
\label{sec:making}
In this section we show that making contact between two parts $A$, $B$, of an economy (forming a partition) never decreases the sum of $\log Z$ (at extensive order).  We will prove this first in the basic case where the two parts are initially not in contact at all and making contact allows exchange of all types of good and where each agent's utility depends only on its own amounts of possessions.  

Denote the initial endowments by $P^0_A,P^0_B$, and let $P = P^0_A+P^0_B$.
Then the partition function for the joint system is
$$\tilde{Z}(P) = \int \prod_{i \in A\cup B} u_i \, \delta(\sum_{i\in A\cup B} p_i - P) \prod_{i \in A\cup B} dp_i.$$
Introducing $P_B = \sum_{i \in B} p_i$ and the partition functions for the two parts from (\ref{eq:Z}), this can be written as
$$\tilde{Z}(P) = \int Z_A(P-P_B) Z_B(P_B)\, dP_B.$$

The extensivity assumption (\ref{eq:zeta}) is that for each economy, $\log Z(P) = N \zeta(P/N)$ for a differentiable function $\zeta$.  It follows that for an order $1$ change $\delta P$ in one component of $P$, the change in $\log Z$ is $\zeta'(P/N) \delta P$ to leading order, with $\zeta'$ being the derivative for the given component.  Thus the change in $Z$ is by a factor $\exp(\zeta'(P/N) \delta P)$ to leading order.  This factor is at least $\frac{1}{e}$ if $\delta P \le 1/|\zeta'|$ (capped at a constant if $\zeta'$ is small).  Extending to $L$ types of good, $Z$ decreases by a factor at most $\frac{1}{e}$ for $\delta P$ in a volume of order $\prod_n \frac{\partial \zeta}{\partial p_n}^{-1}$.

By the above result, there is a neighbourhood of $P^0_B$ with an order $1$ volume $V$ in which $Z_A(P-P_B)Z_B(P_B) \ge \frac{1}{e^2} Z_A(P^0_A) Z_B(P^0_B)$.   Then
$$\tilde{Z}(P) \ge \frac{V}{e^2} Z_A(P^0_A) Z_B(P^0_B).$$
It follows that on the extensive scale, $\log \tilde{Z}(P^0_A+P^0_B) \ge \log Z_A(P^0_A) + \log Z_B(P^0_B)$, as claimed.

A more general case is that $A$ and $B$ already have contact for some types of good and the trader just adds contact for some more types, but this can be handled the same way.

A further generalisation is to allow agents' utility functions to depend on the amounts of other agents' possessions too.  Then the expression for the utility function of agent $i$ may change when two parts are put into contact.  Before the contact, the utility depends only on the possessions of agents in the same part as $i$, whereas after the contact there may be a dependence on the possessons of agents in the other part too.  

We take care of this case by making a ``summability'' assumption.  
The idea is the same as for the theory of Gibbs states in statistical mechanics \cite{Ge}. We write $u_i = \exp \phi_i$ and so the product of utilities becomes the exponential of the sum of the $\phi_i$.  The assumption is that for a partition of the set of agents into two parts, $A$ and $B$, the effect of turning on the dependence of $\sum_{i\in A} \phi_i$ on the possessions of the agents in $B$ is of order 1 with respect to $N$.  Then the above arguments go through in the thermodynamic limit.  This can be a reasonable assumption.  For some agents in $A$, say those near the frontier with $B$, the effect on $\phi_i$ can be significant, but for most of them it is assumed to be small so that even though there are order $N$ agents in $A$, the total effect is still bounded.

\section{Breaking contacts in a system}
\label{sec:breaking}
We first treat the case where each agent's utility is independent of the possessions of the other agents.
From the previous section, the marginal density for $P_B$ in a joint system $A \cup B$ is $$\rho(P_B) = Z_A(P-P_B) Z_B(P_B)/\tilde{Z}(P).$$
The probability that $P_B$ is in the region where $\rho$ is exponentially small is exponentially small.  To see this, the volume in $P_B$-space is finite and scales like $N^L$: $V=\prod_n P_n$, so for any $\rho_0 \in \R$, $\int_{\{\rho \le \rho_0\}} \rho\, dP_B \le \rho_0 V$.  
So on splitting the system, with all but exponentially small probability, $P_B$ will be at a value such that on the extensive scale, $\rho \sim 1$, so $\log Z_A + \log Z_B = \log\tilde{Z}$, i.e.~there is no change in the total $\log Z$.

Cases where an agent's utility may depend on other agents' possessions can be handled similarly to the previous section, subject to the summability assumption given there.

\section{Trading contact with trader}
\label{sec:trade}

To keep description simple, we suppose the trader offers a price vector $\mu_T$ covering all types of good (other than money) in the target economy or part-economy, with all components positive.  There are obvious modifications for the case of a proper subset of types of good.  Denote the initial money and (vector of) goods in the economy by $M,G$, respectively.

In trading contact at price vector $\mu_T$, the dynamics is again reversible but with respect to a new stationary density 
$$\frac{1}{\tilde{Z}(M_0,\mu_T)} \prod_i u_i\, \delta\left(\sum_i m_i + \mu_T \cdot \sum_i g_i - M_0\right),$$ with $M_0 = M+\mu_T \cdot G$ and partition function 
$$\tilde{Z}(M_0,\mu_T)=\int  \prod_i u_i\, \delta\left(\sum_i m_i + \mu_T \cdot \sum_i g_i - M_0\right) \, \prod_i dm_i dg_i.$$  
Writing vector $\tilde{G} = \sum_i g_i$, and the concomitant amount of money $\tilde{M} = M_0 - \mu_T \cdot \tilde{G}$,
the marginal density for $\tilde{G}$ is 
$$\rho(\tilde{G}) = Z(\tilde{M},\tilde{G})/\tilde{Z}(M_0,\mu_T).$$  
By extensivity (following similar lines to section~\ref{sec:making}), in an order 1 volume $U$ around the initial point $G$, $Z(\tilde{M},\tilde{G}) \ge \frac{1}{e} Z(M,G)$, so $\tilde{Z} \ge \frac{U}{e} 
Z(M,G)$.  Thus $\log \tilde{Z} \ge \log Z(M,G)$ on the extensive scale.

On breaking the contact, $\tilde{G}$ is chosen with probability density $\rho$.  It is exponentially unlikely for $\rho$ to be exponentially small.  Thus $\log Z(\tilde{M},\tilde{G}) = \log \tilde{Z}$ on the extensive scale, and we deduce that $\log Z(\tilde{M},\tilde{G}) \ge \log Z(M,G)$.

\section{Checking the axioms}
\label{sec:checking}
The thermal macroeconomic theory of \cite{CM} is based on a string of axioms, following \cite{LY}.  Here we check that they hold for the thermodynamic limit of the class of exchange economies under consideration.

Axiom A0 is that each economic system with specified values of conserved quantities goes to a unique statistical state.  This holds for a class of such systems with all-to-all encounters \cite{M25} and the result is expected to generalise.

Axioms A1,A2,A3, and also A6,A11,A12,A13' are automatic, so we do not list them here.

Axiom A4 says that for each system $A$ with state $X$, and $\lambda>0$, one can consider a scaled version $\lambda A$ with scaled state $\lambda X$, and if $X\preceq Y$ for $A$ then $\lambda X \preceq \lambda Y$ for $\lambda A$.  This holds by the extensivity assumptions in section~\ref{sec:strategy}.  

Axiom A5 says that for any $\lambda \in (0,1)$ any system $A$ can be subdivided into two unconnected parts $\lambda A, (1-\lambda)A$, and the state $X$ of $A$ is reversibly accessible from $(\lambda X, (1-\lambda)X)$.  This is straightforward, from sections~\ref{sec:making} and \ref{sec:breaking}.

Axiom A7 says that $(\lambda X, (1-\lambda)Y) \preceq \lambda X + (1-\lambda) Y$.  This follows from Section~\ref{sec:making}.

Axiom A8 says that for all $M>0$ and states $X$ then $X \prec X+M$, where $X+M$ denotes the state where the money component of $X$ is increased by $M$.  Section~\ref{sec:fin} showed that $X \preceq X+M$.  So it remains to show that $X+M \not\preceq X$.  The trader offering goods at a price could reduce the money in $X$ but only with an associated change in amount of some other good (sec.~\ref{sec:trade}).  Making or breaking contacts (sections \ref{sec:making}, \ref{sec:breaking}) doesn't change the money in the system.  Using an external system, the trader could reduce the money in $X$ but a given reduction cannot be achieved by an arbitrarily small change in the external system.  These are the only ways we allow the trader to interact with the system.  So $X+M \not\preceq X$.

Axiom A9 says that the {\em accessible region} $A_X = \{Y: X \preceq Y\}$ from a state $X$ has a unique support plane at $X$ and it varies Lipschitz continuously with $X$.  To justify this, we first show the following characterisation of $A_X$.
\vskip 1ex
\noindent {\bf Lemma 1}: In the thermodynamic limit, $A_X = \{Y:~\log Z(Y) \ge \log Z(X)\}$.
\vskip 1ex
\noindent{\bf Proof}: From Sections~\ref{sec:fin}--\ref{sec:trade}, all four ways the trader can act on a system in the thermodynamic limit result in $\log Z(Y) \ge \log Z(X)$.  So what we have to show is that for any state $Y$ with $\log Z(Y) \ge \log Z(X)$, there is a way the trader can act to move $X$ to $Y$.  Because the trader is allowed to make small changes to an external system, it is enough to prove this for $\log Z(Y) > \log Z(X)$.  First we show that $\log Z$ is concave.  Given states $X_0, X_1$ and $\lambda \in (0,1)$, the state $U=(1-\lambda)X_0,\lambda X_1)$ of two unconnected scaled copies has $\log Z(U) = (1-\lambda) \log Z(X_0) + \lambda \log Z(X_1)$ because of the scaling assumption and $Z$ for an unconnected pair of systems is the product.  On putting the two systems into contact, the state goes to $V = (1-\lambda)X_0+\lambda X_1$, and by Section~\ref{sec:making}, $\log Z(V) \ge \log Z(U)$.  So $\log Z$ is concave.  It follows that the super-level sets of $\log Z$ are convex.  Writing $M$ for the money component of state and vector $G$ for the amounts of remaining goods, let $\beta = \frac{\partial}{\partial M} \log Z(Y)$ and $\nu = \frac{\partial}{\partial G} \log Z(Y)$.  Then by concavity, on the plane $\beta (M-M_Y) + \nu (G-G_Y)$, $\log Z$ achieves its maximum at $Y$.  Using the assumption that all components of $\nu$ are positive, 
the trader can move any state on this plane to $Y$ by offering to trade at the price vector $\mu = \nu/\beta$. If $X$ is below this plane then the trader can move $X$ up to the plane by adding money, and hence to $Y$.
If $X$ is above the plane then one can make a finite chain of such steps between intermediate states from $X$ to $Y$.  We skip the details.  
  \qed
\vskip1ex
\noindent $A_X$ is convex by A7.  By the representation in Lemma 1, it has unique support plane given by the set of $Y$ such that $\nu^T (Y-X) = 0$, where $\nu_t =\frac{\partial}{\partial Y_t} \log Z(Y)$ at $Y=X$ for types $t$ of good.  Because $\zeta$ was assumed to be $C^{1+Lip}$, $\nu$ is Lipschitz continuous in $X$ and hence this plane varies Lipschitz continuously with $X$.

Axiom A10 says that the boundary $\partial A_X$ of $A_X$ is connected.  $A_X$ is a closed convex subset of $\R^L$. For such a set, if its boundary is not connected then $A_X$ is the slab between two parallel hyperplanes \cite{Fi}.  But on adding an arbitrary positive amount of money to any state in $A_X$, it moves to the interior of $A_X$ (A8).  This can't be true simultaneously for points on the two hyperplanes, giving a contradiction.

Axiom A13 says that ``financial equilibrium'' is transitive.  Systems $A$ and $B$ with states $X,Y$ are said to be in {\em financial equilibrium}, denoted $X \equiv Y$, if on putting them in financial contact there is no nett money flow from one to the other.  Denote the amounts of money in the state of $A$ by $M_A$ and similarly for $B$ and suppress mention of the amounts of other goods in $A$ and $B$, as they are unaffected by financial contact.  Financial contact is a special case of section~\ref{sec:making}.  Thus
\begin{equation}
\log Z_A(M_A) + \log Z_B(M_B) \le \log \tilde{Z}(M),
\label{eq:A13}
\end{equation}
where $\tilde{Z}$ is for their financial join and $M=M_A+M_B$.  By Section~\ref{sec:breaking} there is no nett money flow iff equality holds in (\ref{eq:A13}).  
\vskip 1ex
\noindent {\bf Lemma 2}: For two systems put into financial contact, there is no nett money flow iff they have the same {\em coolness} $\beta = \frac{\partial} {\partial M}\log Z$.
\vskip 1ex
\noindent{\bf Proof}: If there is no nett money flow then we have equality in (\ref{eq:A13}), so it follows that $$f(x) = \log Z_A(x) + \log Z_B(M-x)$$ is maximised over $x$ at $x=M_A$.  Then by differentiation with respect to $x$ we deduce that they have equal coolness, $\beta_A = \beta_B$.
For the converse, recall that $\tilde{Z}(M) = \int Z_A(M_A) Z_B(M-M_B)\, dM_A$, so to obtain equality in (\ref{eq:A13}) (in the thermodynamic limit) we need $M_A$ to be at a non-degenerate maximum of $f$.
We assumed in Section~\ref{sec:strategy} that for any system, $\beta$ has negative derivative with respect to mean money per agent, so
$$\frac{\partial}{\partial x} (\beta_A(x) - \beta_B(M-x)) <0.$$  So if $\beta_A(M_A) = \beta_B(M-M_A)$ then $M_A$ is a non-degenerate maximum of $f$, hence there is no nett money flow. 
Combining the two directions, $A$ and $B$ are in financial equilibrium iff $\beta_A = \beta_B$.  \qed
\vskip1ex
\noindent Then transitivity of financial equilibrium is trivial ($\beta_A=\beta_B$ and $\beta_B = \beta_C$ implies $\beta_A = \beta_C$).

Our justification of axiom A14 uses that of A15, so we treat A15 first.

Axiom A15 says that for every pair $A,B$ of simple systems and states $X$ of $A$ and $Y$ of $B$ there exists $M>0$ such that either $X \equiv Y+M$ or $Y \equiv X+M$.   
Given state $X$ of $A$, take any state $Y_0$ of $B$ and consider the line of states of $B$ formed by adding or removing money from $Y_0$.  By the assumption about $\beta$ going continuously from $+\infty$ to $0$, there is a point $Y$ on this line at which $\beta$ is the same as for $X$.  
Then by Lemma 2, they are in financial equilibrium.
%Putting the two in financial contact produces a joint system with marginal density for the amount $M_A$ of money in $A$ being $$Z_A(M_A) Z_B(M-M_A)/\tilde{Z}(M),$$ with $M$ the total amount of money.  By the equality of $\beta$ for the two systems and the assumption $\frac{\partial \beta}{\partial M}<0$, this probability density is peaked around the initial allocation.  Using the extensivity assumption, the probability of $M_A$ being outside any extensive neighbourhood of the initial allocation is exponentially small.  
So A15 holds.

Finally, axiom A14 says that for any state $X$ of a simple system $A$ there are states $X_0 \equiv X_1$  such that $X_0 \prec X \prec X_1$.  We assume that the economy has at least one other type of good besides money (else the axiom can not in general be satisfied) and that
for this good the assumption about $\nu/\beta$ from Section~\ref{sec:strategy} holds.  To interpret two states of the same system being in financial equilibrium, we have to clone two copies $A_0,A_1$ of the system; states $X_0, X_1$ are in financial equilibrium if there is no nett money flow on putting them in financial contact.  As in the treatment of A13 above, financial equilibrium is equivalent to equal coolnesses.
Our strategy to prove A14 for our systems is firstly to let $M$ be any positive amount of money less than that in the initial state $X$ and 
let $X_1 = X+M$ and $X_0' = X-M$.  By $\frac{\partial \beta}{\partial m}<0$, $X_0'$ has larger $\beta$ than $X$, which in turn has larger $\beta$ than $X_1$.  Then, starting from $X_0'$, sell goods near market price, making a state $X_0$ preserving $X_0 \prec X$, until $X_0$ reaches the same $\beta$ as $X_1$.  Selling near market price implies that along this path, $dM$ is only a little less than $-\frac{\nu}{\beta} dG$.  Putting this into
$$d\beta = \frac{\partial \beta}{\partial M} dM + \frac{\partial \beta}{\partial G} dG = \frac{\partial \beta}{\partial M} dM + \frac{\partial\nu}{\partial M} dG$$ (using symmetry of second partial derivatives) and using the assumption that $\beta/\nu$ decreases from $+\infty$ to $0$ as $m$ increases, we obtain that $\beta$ can be decreased arbitrarily and hence can be made equal to that for $X_1$, so we use the resulting state for $X_0$.  

%consider the states $X_{-M_T}$ and $X_{+M_T}$.  By section~\ref{sec:fin}, $X_{-M_T} \prec X \prec X_{+M_T}$.  Then we let the trader trade equal and opposite amounts of goods with the two copies near their respective market prices, thereby making negligible changes in $\log Z$ for each.  We have to justify that this is possible.  The idea is that at the minimum of $\log \tilde{Z}(M+\mu_T\cdot G,\mu_T)$ (from section~\ref{sec:trade}) with respect to $\mu_T$, $\log{\tilde{Z}} = \log Z(M,G)$.  For suitable nearby $\mu_T$, the trader moves goods from one copy to the other, taking or losing some money in the process because of the difference in prices.  Our hope is that there is a direction in this space of transfers that leads to $\beta$ in the two parts becoming equal.  NOT OBVIOUS!  Requires an additional awkward assumption!

\section{Completing the proof of the formula for entropy}
\label{sec:entropy}
In each mode of interaction with the trader, we have shown that $\log Z$ never decreases by an extensive amount.  Thus the trader can never move a collection of economies to a state with lower total $\log Z$.
In the other direction, we have proved in Section~\ref{sec:fin} that the trader can increase $\log Z$ for an economy by any positive amount, simply by making money available to it.  These results are not enough, however, to identify $\log Z$ as an entropy function for a system.  It is essential to show it has the required scaling property.

So we follow the construction of \cite{LY}.  Given two states $X_0\prec X_1$ and arbitrary $\lambda_0, \lambda_1\ge 0$ with $\lambda_0+\lambda_1 = 1$, we find the set of states $X$ reversibly accessible (meaning accessible in both directions) from $(\lambda_0 X_0, \lambda_1 X_1)$ (meaning unconnected scaled copies).  For each $X$ with $X_0 \prec X \prec X_1$ there is a unique $\lambda_1$ such that this holds.  Then \cite{LY} prove that $\lambda_1(X)$ is an entropy function.

Put the two parts into contact via an ``exchange line'' $\lambda_0 (\tilde{X}_0 -X_0) = t \xi$, for some $\xi \in \R^L$, parametrised by $t \in \R$, and then disconnect them to achieve a state $(\lambda_0 \tilde{X}_0, \lambda_1 \tilde{X}_1)$.   Then $\lambda_1 (\tilde{X}_1 - X_1) = -t \xi$.  By Section~\ref{sec:making}, the system goes to an equilibrium $t^* \ne 0$ with higher $\log Z$ or stays at $t^*=0$.  
Define $$\tilde{S}(\tilde{X}_0) = \lambda_0 \log Z(\tilde{X}_0) + \lambda_1 \log Z(\tilde{X}_1)$$ in the space where $\lambda_0 \tilde{X}_0 + \lambda_1 \tilde{X}_1$ is constant.
The case $t^*=0$ occurs iff $\xi$ is tangent to the curve $\tilde{S}=$ constant.
We want to make the resulting $\tilde{X}_0, \tilde{X}_1$ nearly equal, then we can merge them nearly reversibly.

To get arbitrarily close to any point $\tilde{X}_0$ where
$\tilde{S}(\tilde{X}_0) = \tilde{S}(X_0)$ and $\lambda_0 \tilde{X}_0 + \lambda_1 \tilde{X}_1 = \lambda_0 X_0 + \lambda_1 X_1$, choose a differentiable curve $\gamma$ from $X_0$ to near $\tilde{X}_0$ along which $\tilde{S}$ is strictly increasing (this exists by convexity).  The path $\gamma$ can be approximated to arbitrary precision by a finite sequence of linear paths, each piece of which ends at an equilibrium as above.  We can get back from $\tilde{X}_0$ to arbitrarily close to $X_0$ by the same procedure.
So the set of $(\tilde{X}_0, \tilde{X}_1)$ that is reversibly accessible from $(X_0,X_1)$ contains the level set of $\tilde{S}$ through the latter.
The subset where $(\lambda_0,\tilde{X}_0, \lambda_1 \tilde{X}_1)$ can be reversibly merged is where $\tilde{X}_0 = \tilde{X}_1$ (by Section~\ref{sec:breaking}).

Using the asymptotic scaling assumption, this means that the set of $X$ reversibly accessible from $(\lambda_0 X_0, \lambda_1 X_1)$ is contained in the set where $\log Z(X) = \lambda_0 \tilde{S}_0 + \lambda_1 \tilde{S}_1$, where $\tilde{S}_j$ denote the values of $\tilde{S}$ at $X_j$.  Now $\tilde{S}_1 > \tilde{S}_0$ because $X_0 \prec X_1$.  Use $\log Z(X_0) < \log Z(X_1)$ from Sections~\ref{sec:fin}--\ref{sec:trade}.  So
$$\lambda_1(X) = \frac{\log Z(X)-\tilde{S}_0}{\tilde{S}_1 - \tilde{S}_0}.$$
This is an orientation-preserving affine transformation of $\log Z$, so without loss of generality we can take the entropy function to be $\log Z$.

It remains to check that $\log Z$ is calibrated, that is, there is no need to scale it by different constants for different systems in order to obtain their total entropy.  The reason is that for the unconnected product of two systems, the probability density is the product of the probability densities, so the partition function $Z(P_1,P_2) = Z_1(P_1)Z_2(P_2)$.  Thus $\log Z = \log Z_1 + \log Z_2$, which is the condition required for calibrated entropy functions.

\section{Comments}
\label{sec:comments}
The paper makes concrete the idea presented in \cite{CM} that economic systems maximise liberty at the aggregate level.  Here, liberty is interpreted as the accessible volume in the space of micro-states for given macro-state, quantified by the logarithm of the partition function.

The paper could also be viewed as computation of the entropy function for a class of microscopic models of physical thermodynamic systems in the programme of \cite{LY}, which we have not seen done.  It is possible that the results are all well known using older formulations of thermodynamics, such as Carath\'eodory's axioms (or Giles' version \cite{G}).  More formal treatments could probably be made along the lines of \cite{E}.

The utilities of each agent could be multiplied by any positive factor $\gamma$ without changing the dynamics.  This changes $\log Z$ by adding $N \log \gamma$.  But the zero of entropy for an economy has no significance in our theory (though this would change if we allow migration of agents).  In contrast, the scale for entropy has an intrinsic meaning for these micro-economic models, so it is natural to set the scale for entropy using them (compare setting the temperature scale by using an ideal gas in physics, and hence the entropy scale).

Similarly the encounter matrix $k$ could be multiplied by a factor, which just speeds up the process by that factor and makes no difference to $\log Z$.

%Note possible resolution of Gibbs' paradox via gradations of substitutes. EXPAND

A question is how to handle mean-field models, in which the utility for an agent depends on its own possessions and the mean possessions of all the others,
e.g.~the ``Bouchaud'' economies of \cite{LMC}.  The relation $S=\log Z$ was used in Appendix D to \cite{LMC}, but it is not clear that it follows from the analysis of the present paper, because the summability condition fails.  This must be a known issue in statistical mechanics.

\section*{Acknowledgements}
I am grateful to Nick Chater and Daniel Ueltschi for discussions.

\section*{Appendix:~Relation to canonical partition function}
For some purposes it is convenient to weight the amounts of each type of good exponentially instead of constraining them to given totals. Thus we define the ``canonical partition function''
$$Z_c(\nu) = \int \prod_i u_i \exp(-\nu^T p) \prod_i dp_i$$
for a covector $\nu^T$ with all components positive.  It is immediate that
$$Z_c(\nu)  = \int Z(P) \exp(-\nu^T P)\, dP .$$
We have seen that we can write $Z = \exp S$, where $S$ is the entropy.  Then $Z_c$ is dominated by a neighbourhood of a maximum of $S(P)-\nu^TP$ over $P$.  Let us denote a point of maximum by $P^*(\nu)$.  $P^*(\nu)$ might be on the boundary of allowable $P$ or in the interior.  In either case, for $N$ large the integral can be approximated by $A \exp(S(P^*)-\nu^TP^*)$ with $A$ a prefactor independent of the system size (or weakly dependent on $N$).  Thus on the scale of system size we obtain
$$\log Z_c(\nu) = \max_P (S(P)-\nu^TP).$$

It is convenient to define the {\em free energy} $$F(\nu) = -\log Z_c(\nu)$$
(in Physics, the free energy is $-\frac{1}{\beta}\log Z_c$, considering money to play the role of energy);
then this becomes a standard Legendre transform:
$$F(\nu) = \min_P (\nu^T P - S(P)).$$
Under the assumption that $S$ is concave (which is a consequence of the axioms of \cite{LY}), this Legendre transform can be inverted to deduce that
$$S(P) = \min_{\nu}\,  (\nu^T P -F(\nu)),$$
which is sometimes a convenient way to compute the entropy $S$.  

For example, if the population contains agents who treat goods and money as perfect substitutes, say $u_i(m,g) = (m+g)^{\alpha-1}$, then Appendix C in \cite{LMC} computes the free energy $F$ and the equilibrium amounts $M$ and $G$ as functions of their values $\beta$ and $\nu$.  Then after some cancellation, we deduce that $$S(M,G) = N(\alpha+1) -F(\beta,\nu),$$ which gives the entropy in parametric form (to make it an explicit function of $M,G$ requires an explicit inversion of the mapping from $(\beta,\nu)$ to $(M,G)$).

We note in passing that analysis of substitutes might be a way to resolve the mixing paradox, that the entropy of two distinguishable gases increases on mixing them, whereas it does not if they are indistinguishable.

As another example, we treat here the case of
an economy for whose agents two types of good are complements (meaning they have utility only when matched).  
Concretely, take $u_i(g,m) = \min(m,g)^{\alpha-1}$.  Then we obtain
$$F(\beta,\nu) = N \left( (\alpha-1) \log (\beta+\nu) + \log \beta + \log \nu\right)$$
Hence $$M = \frac{\partial F}{\partial\beta} = N \left(\frac{\alpha-1}{\beta+\nu} + \frac{1}{\beta}\right), \quad G = \frac{\partial F}{\partial \nu} = N \left(\frac{\alpha-1}{\beta+\nu} + \frac{1}{\nu}\right),$$
and again
$$S(M,G) = N(\alpha+1) - F(\beta,\nu),$$
but with the $F$ for complements.


\begin{thebibliography}{WWW}
\bibitem[CM]{CM} Chater NJ, MacKay RS, Thermal macroeconomics, arxiv:2412.00886 
\bibitem[E]{E} Ellis RS, Entropy, large deviations and statistical mechanics (Springer, 1985).
\bibitem[Fi]{Fi} Fischer D, stackexchange 3776710 (2020)
\bibitem[Ge]{Ge} Georgii H-O, Gibbs measures and phase transitions (de Gruyter, 2011).
\bibitem[Gi]{G} Giles R, Mathematical foundations of thermodynamics (Pergamon, 1964).
\bibitem[GG]{GG} Greenberg M, Gao HO, 25 years of random asset exchange models, Eur Phys J B 97 (2024) Article number 69.
\bibitem[LY]{LY} Lieb E, Yngvason J, The mathematical structure of the second law of thermodynamics, in:~Current developments in mathematics 2001 (International Press, 2002), 89--130.
\bibitem[LMC]{LMC} Luo Y, MacKay RS, Chater NJ, Tests of thermal macroeconomic theory on simulated micro-economies, J Phys Complexity 6 (2025) 015008
%\bibitem[M11]{M} MacKay RS, Robustness of Markov processes on large networks, J Difference Eqns \& Applns 17 (2011) 1155--67.
\bibitem[M25]{M25} MacKay RS, Convergence to equilibrium for a class of exchange economies, arxiv:2506.11770, subm to Stochastics (2025)
\bibitem[TSB]{TSB} Toscani G, Sen P, Biswas S, Kinetic exchange models of societies and economies, Phil Trans Roy Soc A 380 (2022) 20210170
\end{thebibliography}
\end{document}